\documentclass{article}

\usepackage{PRIMEarxiv}

\usepackage[utf8]{inputenc} % allow utf-8 input
\usepackage[T1]{fontenc}    % use 8-bit T1 fonts
\usepackage{hyperref}       % hyperlinks
\usepackage{url}            % simple URL typesetting
\usepackage{booktabs}       % professional-quality tables
\usepackage{amsfonts}       % blackboard math symbols
\usepackage{nicefrac}       % compact symbols for 1/2, etc.
\usepackage{microtype}      % microtypography
\usepackage{lipsum}
\usepackage{fancyhdr}       % header
\usepackage{graphicx}       % graphics
\graphicspath{{media/}}     % organize your images and other figures under media/ folder
\usepackage{multirow}

\title{Leveraging Sustainable Systematic Literature Reviews 
}

\author{
  Vinicius dos Santos \\
  University of São Paulo \\
  São Carlos, Brazil \\
  \texttt{vinicius.dos.santos@usp.br} \\
  \And
  Rick Kazman \\
  University of Hawaii \\
  Honolulu, USA \\
  \texttt{kazman@hawaii.edu} \\
  \And
  Elisa Yumi Nakagawa \\
  University of São Paulo \\
  São Carlos, Brazil \\
  \texttt{elisa@icmc.usp.br} \\
}

\begin{document}
\maketitle

\begin{abstract}
Systematic Literature Reviews (SLRs) are a widely employed research method in software engineering. However, there are several problems with SLRs, including the enormous time and effort to conduct them and the lack of obvious impacts of SLR results on software engineering practices and industry projects. To address these problems, the concepts of \textit{sustainability} and \textit{sustainable SLR} have been proposed, aiming to raise awareness among researchers about the importance of dealing with SLR problems in a consistent way; however, practical and concrete actions are still lacking. This paper presents concrete directions towards sustainable SLRs.  We first identified 18 ``green drivers'' (GD) that could directly impact SLR sustainability, and we distilled 25 sustainability indicators (SI) associated with the GD to assess SLRs regarding their sustainability. A preliminary evaluation was conducted on the ten top-cited SLRs in software engineering published over the last decade. From this analysis, we synthesized our insights into 12 leverage points for sustainability. Our results indicate that even in high-quality reviews, there are threats to sustainability, such as: flaws in the search process, lack of essential details in the documentation, weak collaboration with stakeholders, poor knowledge management, lack of use of supporting tools, and a dearth of practical insights for software engineering practitioners. The good news is that moving towards sustainable SLRs only requires some simple actions, which can pave the way for a profound change in the software engineering community's mindset about how to create and sustain SLRs.
\end{abstract}

% keywords can be removed

\section{Introduction}

%% SLR
The practice of Systematic Literature Reviews (SLRs) was first introduced in software engineering to support researchers in aggregating results from empirical studies and to aid the decision-making process in the software industry ~\cite{Kitchenham04}. The SLR method is widely known and employs a well-defined 3-phase process, namely, planning, conduction, and reporting~\cite{Kitchenham15}. The SLR method offers several advantages and also some drawbacks~\cite{Felizardo2020academicContext}. For researchers, the benefits include the improvement of research skills \cite{Pejcinovic15, Borrego15} and the possibility of learning from prior work  \cite{Babar09, Zhang12}. For the software engineering community, the benefits include: the ability to dealing with information from different studies in an unbiased manner \cite{Kitchenham15,Niazi15},  the production of auditable and repeatable results \cite{Kitchenham2011Repeatability,Budgen2018Reporting}, and the identification of research gaps and perspectives for future work \cite{Kitchenham15,Babar09}. The drawbacks are often associated with human factors, such as the lack of adherence to well-established guidelines \cite{Kuhrmann2017}, poor SLR documentation~\cite{Zhou2016Threats,Ampatzoglou2019}, or even lack of specialists to support critical activities~\cite{Carver2013}. Other drawbacks are caused by technical factors such as the lack of maturity of supporting tools, thus requiring a lot of manual work~\cite{AlZubidy2018}. Despite some solutions having been proposed, such as the use of iterative processes~\cite{Fabbri2013} or specific supporting tools \cite{Felizardo2020Automating}, these solutions are incapable of addressing the SLR problems as a whole. 

The concept of \textit{sustainability} was introduced to the field of SLRs as a disruptive vision about how to deal with SLR problems in a holistic way~\cite{Santos2021}. \textit{Sustainability in SLRs} refers to a process and a set of actions to allow SLRs to endure over time (i.e., longevity) with less time and effort consumed and an improved impact on industry~\cite{Santos2021}. 
Sustainability in SLRs involves three dimensions: (i) the social dimension addresses human aspects; (ii) the economic dimension is concerned with the resources associated with SLR conduction and update (mainly the time and effort of stakeholders); and (iii) technical dimension preserves the reliability of supporting tools used to conduct SLR. To address these challenges, the concept of a \textit{sustainable SLR} was introduced, defining a set of core characteristics and critical factors\cite{santos2024}. The characteristics highlight the importance of providing reliable results and rich documentation, employing a process that leverages best practices, and adopting techniques such as interactivity and pilot testing. Sustainability also involves thinking about the long-term view: using resources responsibly,  maintaining SLRs continuously, and avoiding waste in the research process. The third goal of sustainability addresses the usefulness of the results and the importance of providing evidence for a broad community, including academic communities and software practitioners. 

The state of the practice of SLRs in software engineering presents several problems. It appears that promoting the concept of a sustainable SLR could be a solution, but there is currently a lack of practical and concrete actions to achieve more sustainable SLRs. Hence, our main research question is: \textit{What changes should be made to achieve more sustainable SLRs?}

The goal of this paper is to present concrete directions towards sustainable SLRs, including recommendations for software engineering researchers about near-term actions that should be taken. To this end we first identified a set of 18 ``green drivers'' (GD) that can directly impact SLR sustainability. Next, we applied the Goal-Question-Metric \cite{vanSolingen2002} method to identify a set of sustainability indicators (SI). We selected the ten top-cited software engineering SLRs published in the last decade and carefully analyzed those reviews from social, economic, and technical perspectives using the indicators. We found various flaws in those studies, from the sustainability perspective, that helped us refine our recommendations. Finally, we propose 12  practical recommendations to leverage SLR sustainability. The core contribution of this work is to introduce a \textit{sustainability-aware} mindset to the software engineering community conducting SLRs and, ultimately, to address the existing problems with SLRs.

The rest of the paper is organized as follows: Section~\ref{sec:ourIdea} shows how we obtained the GD and SI; Section~\ref{sec:evaluation} presents the preliminary evaluation and its results; Section~\ref{sec:recommendations} presents the practical recommendations; and Section~\ref{sec:futurePlans} depicts the future plans. 

\section{Green Drivers and Sustainability Indicators}
\label{sec:ourIdea}

To find practical recommendations to leverage sustainable SLRs, we first needed to derive the GD and define the SI. The GD refers to points that deserve attention to leverage the sustainability of SLRs, so researchers can consider them when aiming for sustainable SLRs. SI make it possible to measure SLRs from a sustainability perspective. This section summarizes our process\footnote{The external material details this process: \url{https://anonymous.4open.science/r/NIER-LvPSSLR-E0F6}} that contains two main steps: identification of GD and identification of SI based on GD.

As our first step, we analyzed the 15 characteristics and 15 critical factors proposed by Santos et al. \cite{santos2024} to identify the GD that could directly impact social, economic, or technical aspects of sustainable SLRs. For this, we applied thematic analysis \cite{cruzes2011} using open coding to identify key codes representing those GD; next, those codes were clustered to generate a unique set of GD. For instance, a characteristic of sustainable SLRs is that \textit{SLRs should be produced with responsible use of resources (e.g., time, human effort, and monetary cost) and aiming at reducing time consumption}. Two critical factors also mention that sustainable SLRs should present an \textit{efficient management/use of resources} and \textit{use of techniques that minimize resource consumption}. Hence, we synthesized this characteristic and factors into a single GD (GD5 - \textit{Resources use}). This step resulted in 18 GD, as shown in Table \ref{tab:sustainabilityIndicators}.

In the second step, we used the GD to derive the set of SI; for this, we applied a GQM template. For example, considering GD5 (\textit{Resources use}) and using the GQM template, we had the following: \textit{Analyze} the resources use \textit{for the purpose of} understanding  \textit{with respect to} time/effort waste \textit{from the viewpoint of} researcher \textit{in the context of} a new SLR conduction. We focused on two major effects on SLRs that are: \textit{excessive time/effort consumption} and \textit{the lack of impact in industry}. Our point of view was always from the researcher's perspective, and the context was scientific research. Some of the most time-consuming activities in SLRs include searching studies in databases, selecting primary studies, extracting data, and assessing  study quality \cite{Carver2013}. Hence, for GD5, we defined the number of studies retrieved from the databases and analyzed in data extraction as a metric (SI7 - \textit{Number of studies retrieved/analyzed}). Also, several techniques can reduce time and effort, for instance, automation~\cite{Felizardo2020Automating}, text classification \cite{Watanabe2020}, or even tools \cite{AlZubidy2018}. Hence, the number of techniques used was another metric (SI8 - \textit{Number of techniques used to minimize time and effort consumption}). It is worth highlighting that a major challenge in proposing metrics for SLR sustainability is to ensure they are suitable to be applied in SLRs, and can highlight flaws that impact sustainability. After applying GQM in each GD and analyzing them, we obtained 25 SI, as listed in Table~\ref{tab:sustainabilityIndicators}.

\begin{table*}[!h]
\label{tab:sustainabilityIndicators}
\footnotesize
\caption{Sustainability Indicators derived from Green Drivers}
\begin{tabular}{cp{5cm}cp{8.2cm}}
\hline
\textbf{GD Id} & %\multicolumn{1}{c}
{\textbf{Green Driver (GD)}} & \textbf{SI Id} & \textbf{Sustainability Indicator (SI)} \\ \hline
GD1 & Compliance with SLR guidelines & SI1 & A reference list indicating the guidelines used to conduct the studies  \\ \hline
\multirow{2}{*}{GD2} & \multirow{2}{*}{Iterativity and pilot testing} & SI2 & A boolean indicating the presence of iterations \\ 
 &  & SI3 & A boolean indicating the use of pilot testing \\ \hline
GD3 & Documentation quality & SI4 & A reference indicating the use of guidelines for documentation \\ \hline
\multirow{2}{*}{GD4} & \multirow{2}{2cm}{Study reliability} & SI5 & Number of actions taken to ensure reliability \\ 
 &  & SI6 & A boolean indicating the conduction of quality assurance assessment. \\ \hline
\multirow{2}{*}{GD5} & \multirow{2}{*}{Resources use} & SI7 & Number of studies reviewed/analyzed. \\ 
 &  & SI8 & Number of techniques used to minimize time/effort consumption. \\ \hline
GD6 & Use of tools to support SLR & SI9 & Number of tools used during the process \\ \hline
\multirow{3}{*}{GD7} & \multirow{3}{*}{Accessibility of support technology} & SI10 & A boolean indicating whether the tool used is available for use \\ 
 &  & SI11 & A qualitative analysis using the following classification: (i) closed tool; (ii) free with restrictions; (iii) free for use (open source) \\ \hline
GD8 & Communication among stakeholders & SI12 & Number of methods used to improve communication among stakeholders \\ \hline
\multirow{2}{*}{GD9} & \multirow{2}{*}{Participation / collaboration of stakeholders} & SI13 & A qualitative analysis on the author profiles using the following classification: (i) Academic team; (ii) Industry; (iii) hybrid \\
 &  & SI14 & Number of stages which stakeholders contributed \\ \hline
GD10 & Knowledge of stakeholders & SI15 & Number of published papers about the topic addressed in SLR \\ \hline
GD11 & Experience of team members in SLR & SI16 & Number of secondary studies conducted \\ \hline
GD12 & Knowledge sharing/transfer & SI17 & Number of Knowledge sharing techniques used \\ \hline
\multirow{2}{*}{GD13} & \multirow{2}{*}{Accessibility of SLR artifacts} & SI18 & A boolean indicating whether the protocol is available \\ 
 &  & SI19 & A boolean indicating whether the study have a replication kit \\ \hline
GD14 & Research waste & SI20 & A boolean indicating whether the authors perform an evaluation of similar studies \\ \hline
GD15 & Continuous update & SI21 & Number of updates \\ \hline
\multirow{2}{*}{GD16} & \multirow{2}{*}{Improvement of SLR reusability} & SI22 & Number of studies that reuse components of the study \\ 
 &  & SI23 & Number of components reused/adapted from previous SLR \\ \hline
GD17 & Research usefulness / impacts & SI24 & A boolean indicating the presence of recommendations for industry practitioners \\ \hline
GD18 &  Research impact / long-term goals & SI25 & Number of citations per year \\ \hline
\end{tabular}
\end{table*}

\section{Preliminary Evaluation} \label{sec:evaluation}

Our evaluation had two objectives: to comprehend SI effectiveness in appraising SLRs regarding sustainability and to diagnose flaws in SLRs that represent leverage points that served as bases to derive the practical recommendations. This section summarizes this evaluation, while the external material details it. 

The evaluation encompassed four main steps: (i) \textit{evaluation planning} (we defined the evaluation goals, restrictions, and strategies to extract information from SLRs); (ii) \textit{SLR selection} (we searched for SLRs published in two journals in software engineering (Information \& Software Technology and Journal of Systems and Software) and found 238 reviews. We selected one most-cited review (in Science Direct) per year (2012 to 2021), totaling 10 SLRs\footnote{The external material presents the 10 SLRs.}); (iii) \textit{SI application} (we first conducted a pilot study followed by the application of the 25 SI (Table \ref{tab:sustainabilityIndicators}) in the 10 SLRs); and (iv) \textit{results synthesis} (we synthesized the results, from which we were able to distill the practical recommendations to achieve sustainable SLRs). Following, the results regarding each GD and measures using the SIs are summarized and briefly discussed.

Concerning \textbf{GD1}, 9 in 10 studies follow the basic guidelines proposed by Kitchenham et al.~\cite{Kitchenham04,Kitchenham07}. Two studies also comply with recommendations and lessons learned from Brereton et al.~\cite{Brereton2007Lessons}, Brocke et al.~\cite{Brocke2009}, Staples and Niazi \cite{Staples2007}, Webster and Watson \cite{Webster2002}, and Wohlin ~\cite{Wohlin2014}. 
Regarding \textbf{GD2}, our analysis showed that iterativity was mentioned in only three SLRs, and most refinements occurred during the initial stages of SLRs; moreover, only two reviews explicitly discussed the use of pilot testing. Regarding the adoption of SLR guidelines (\textbf{GD3}), the majority of reviews followed the guidelines proposed by Kitchenham et al. \cite{Kitchenham04, Kitchenham07, Kitchenham09, Kitchenham15}; other guidelines were also used, such as \cite{Brereton2007Lessons, Brocke2009,Staples2007, Webster2002, Wohlin2014}.  

To appraise the reliability of SLRs (\textbf{GD4}), we evaluated the actions taken to mitigate bias, and most of them prioritized the selection of primary studies; additionally, nine SLRs conducted some kind of quality assessment. Regarding the amount of resources (time and effort) (\textbf{GD5}), we observed that the number of studies selected by manual searches and in electronic databases ranged from 707 to 181,829. The large number of studies caught our attention because most SLRs employed small teams (4 or 5 authors), and the use of automation techniques was not mentioned, indicating that much manual effort was likely consumed; only two studies mentioned techniques to reduce effort consumption. During our analysis, we noticed that only four studies mentioned using supporting tools (\textbf{GD6}). The accessibility of the tools (\textbf{GD7}) varies significantly based on their cost and complexity, including commercial, free-use, or open-source tools. Ultimately, many SLRs preferred more accessible, generic tools like spreadsheets and reference managers despite the availability of more specialized alternatives.

Regarding the techniques to support collaboration among researchers (\textbf{GD8}), the most recurrent one was consensus meetings; nevertheless, no clue about which and how decisions were taken was provided. In some cases, SLRs combined consensus meetings with more pragmatic techniques, such as Kappa Coefficient, Krippendorff Alpha Kr$_{a}$, or data extraction cards. Concerning stakeholders (\textbf{GD9}), all authors were affiliated with universities or research centers, and only one was from university and industry. There is also a poor collaboration with people not directly involved in the SLR (i.e., external stakeholders); authors of only one SLR contacted primary study authors to check data extraction and point out inaccuracies.  
Regarding the knowledge and experience of the authors of SLRs (\textbf{GD10} and \textbf{GD11}), our results revealed that 23\% of all authors can be considered experienced in SLRs. We also noticed that most SLRs (8 out of 10) had at least one experienced author, showing the heterogeneity of teams conducting SLRs. Regarding the knowledge sharing and transfer (\textbf{GD12}), few details were provided about internal knowledge sharing and transfer among researchers. Only consensus meetings were used to exchange information in 7 SLRs. In addition, concerning the accessibility to the SLR artifacts (\textbf{GD13}), we observed the raw data provided by authors is often incomplete and includes only a few artifacts such as the selected studies, but none provided a complete list of excluded studies or other core data, making it more difficult to reproduce, update, or audit the reviews. 

Our analysis regarding the research waste (\textbf{GD14}) concluded that nine reviews verified if an SLR had been already conducted on the same topic; however, a common flaw is a lack of a systematic search for SLRs to ensure that authors have not missed any similar SLR that could be at least partially reused. Regarding the continuous update of SLRs (\textbf{GD15}), we identified only two SLRs that were updated. Nevertheless, our investigation also revealed that the 10 SLRs somehow inspired at least 284 other SLRs. In addition, we identified a lack of reuse of elements of an SLR (\textbf{GD16}); only two reused elements, including the set of keywords and database set, and only one study compares their findings with previous studies to draw more reliable conclusions. Hence, the reuse of elements is still lacking and could be better explored.

The usefulness of SLR results for industry practitioners (\textbf{GD17}) was not an explicit concern in the SLRs. Only one review dedicated a section to providing insights for practitioners. Regarding the research impact and the long-term goals (\textbf{GD18}), the reviews had from 16.5 to 56.6 citations per year, meaning that they are far beyond the cut-off metric (by Garousi and Fernandes \cite{Garousi2016Top100}) and making them useful for their areas. Another important aspect is the time span in which the information remains useful. Overall, the SLRs have broadly remained relevant to the academic community over the years, indicating a long-term impact on the software engineering research community.

\section{Recommendations}
\label{sec:recommendations}

The evaluation helped us to identify SLR flaws that could impact their sustainability. For instance, when analyzing GD5, we observed that most SLRs had a large number of primary studies, and few adopted techniques to reduce effort; hence, a leverage point is to \textit{prioritize the preservation of resources and foster the use of techniques to reduce effort} (LP5). The following summarizes the 12 leverage points (LP), together with the associated GD\footnote{The external material details the steps to derive the recommendations from the evaluation results.}.

\begin{itemize}

\item\textbf{LP1}: 
Improve team knowledge about best practices for SLR conduction and always prioritize the adoption of guidelines defined and validated by the community~(GD1).

\item\textbf{LP2}: 
Design SLRs to evolve iteratively, using pilot testing~(GD2).

\item\textbf{LP3}: 
Provide a detailed SLR report using best documentation standards to ensure its quality~(GD3).

\item\textbf{LP4}: 
Identify, prioritize, and apply a reasonable number of techniques to mitigate threats to the validity~(GD4).

\item\textbf{LP5}: 
Prioritize the preservation of resources and foster the use of techniques to reduce efforts~(GD5).

\item\textbf{LP6}: 
Use tools to support the SLR process to the fullest extent~(GD6) and document the experience to foster the development and improvement of these tools.

\item\textbf{LP7}: Foster participation and collaboration of stakeholders~(GD8) by defining and documenting the roles played by each one, always making use of strategies to improve internal and external communication~(GD9).

\item\textbf{LP8}: Prefer hybrid teams combining experienced stakeholders in the SLR process~(GD9) and stakeholders with knowledge in the topic addressed in the SLR~(GD10).

\item\textbf{LP9}: Ensure that knowledge acquired in the SLR process is shared with interested parties~(GD12) and all artifacts are fully accessible for readers, fostering open science~(GD13).

\item\textbf{LP10}: Make an effort to avoid unreasonably duplicating SLRs so reducing research waste~(GD14). Reuse as many elements as possible from prior SLRs~(GD16).

\item\textbf{LP11}:
Design SLRs to enable their continuous update~(GD15).

\item\textbf{LP12:} Improve the usefulness of SLR results by designing SLRs that cover multiple perspectives that can benefit the software engineering community~(GD17) and have long-term impacts~(GD18).

\end{itemize}

We believe these 12 leverage points are a comprehensive set of actions (i.e., \textit{what to do}) to promote sustainable SLRs. In many cases, there are existing solutions to realize some of these actions (\textit{how to do}), but there is a need for new solutions as well.

\section{Future Plans}
\label{sec:futurePlans}

Answering our initial research question, we observe that achieving sustainable SLRs requires many changes, particularly a shift in the mindset of how to deal with SLRs. This paper intends to call attention to a novel research line of \textit{Sustainability of SLRs}, which could ensure that SLRs continue to benefit the research community and aid the software industry while reducing the current SLR problems. Some of the most urgent areas of future work are:

\begin{itemize}

\item \textbf{Sustainability-aware SLR process:} The SLR process has become bloated, time-consuming, and costly.
For instance, more and more new actions emerge to mitigate threads to validity; while these actions are relevant, they should be proposed thinking about sustainability issues. The process should then evolve considering the trade-off between the reliability and sustainability of SLRs.

\item \textbf{Measurement of sustainability of SLRs}: Measuring sustainability is a challenge due to the diverse aspects to be considered. We can refine the set of SIs to improve the coverage of the GDs and conduct more evaluations applying the SIs in other SLRs. Sustainability should also be continuously measured during the whole SLR life cycle, and measurement results could aid researchers in identifying the impacts of their actions and help in taking mitigation actions.

\item \textbf{Integrated sustainability-aware approaches}: Instead of proposing novel approaches (processes, methods, techniques, tasks) for solving SLRs problem as widely observed, a diagnosis of existing approaches is necessary, and new approaches should target sustainability. Such approaches should also be integrated forming a systemic sustainability-aware framework.

\item \textbf{Sustainability-based supporting tools}: One of the pains of SLR conduction is the huge amounts of manual effort. Automated supporting tools are one of the solutions, but they are often focused on specific tasks or phases; hence, they should be integrated into a tool suite along with external tools such as publications databases like Scopus. 

\item \textbf{Open Science mindset}:
The SLR authors should be aligned to the open science movement\footnote{\url{https://unesdoc.unesco.org/ark:/48223/pf0000383323}}, making SLRs accessible to the entire community (e.g., using SLR repositories) and remembering that SLRs result from collaborative development. All artifacts generated during SLR conduction should also be accessible to ensure reproducibility, auditability, and reuse.

\item \textbf{Awareness of the SLRs conducted}: Researchers should be concerned with the utility and impacts of their SLR on their target audience and community, unlike what is observed nowadays. Researchers should also consider the impact of SLRs regarding social, economic, and technical perspectives. At the same time, measuring the utility and impact of SLRs (and also measuring SLR maturity) is an open research opportunity.

\end{itemize}

We believe that our leverage points can serve as a basis for these and other future work. There is a long path from now to actual sustainable SLRs, so with collaborative actions and engagement, the software engineering community can mature research on the \textit{Sustainability of SLRs}, benefiting us all. 

\section*{Acknowledgment}
This study was supported by FAPESP (2019/23663-1, 2023/00488-5, 2024/00329-7) and CNPq (313245/2021-5).

%Bibliography
\bibliographystyle{unsrt}  
\bibliography{references}

\end{document}